\documentclass[3p,times,twocolumn]{elsarticle}

\usepackage{ecrc}
\usepackage{graphicx}
\usepackage{dcolumn}
\usepackage{bm}
\usepackage{amssymb}

\def\beq{\begin{equation}}
\def\eeq{\end{equation}}

\usepackage[usenames]{color}

\newcommand{\be}{\begin{equation}}
\newcommand{\ee}{\end{equation}}
\newcommand{\bea}{\begin{eqnarray}}
\newcommand{\eea}{\end{eqnarray}}
\newcommand{\nn}{\nonumber}

 \input{epsf}

\volume{00}

\firstpage{1}

\journalname{Nuclear Physics B Proceedings Supplement}

\runauth{}

\jid{nppp}

\jnltitlelogo{Nuclear Physics B Proceedings Supplement}

\usepackage{amssymb}

\usepackage[figuresright]{rotating}

\begin{document}

\begin{frontmatter}



\dochead{}

\title{Drag induced radiative energy loss of semi-hard heavy quarks}
 %
 \author[a]{Raktim Abir}
 \author[b]{Shanshan Cao}
 \author[a]{Abhijit Majumder}
 \author[c]{Guang-You Qin}
 \address[a]{Department of Physics and Astronomy, Wayne State University, 666 W. Hancock St., Detroit, 
 MI 48201, USA,}
 \address[b]{Nuclear Science Division, Lawrence Berkeley National Laboratory,  Berkeley, CA 94720, USA,}
 \address[c]{Institute of Particle Physics and Key Laboratory  of Quark and Lepton Physics (MOE),
 Central  China  Normal  University,  Wuhan,  430079,  China.} 
 %
 %
 %
\begin{abstract}
 We revisited gluon bremsstrahlung off a heavy quark in nuclear matter within higher twist formalism.
 In this work, we demonstrate that, in addition to transverse momentum diffusion parameter ($\hat q$),
 the gluon emission spectrum for a heavy quark  is quite 
 sensitive to $\hat e$, which quantify the amount of \emph{light-cone} drag experienced by a parton. 
 This effect leads to an additional energy loss term for heavy-quarks. 
 From heavy flavor suppression data in ultra-relativistic heavy-ion collisions one can now estimate the
 value of this sub-leading non-perturbative jet transport parameter ($\hat e$) from our results. 
\end{abstract}

\begin{keyword}
   energy loss, drag, heavy-quark, bremsstrahlung radiation.
 
\end{keyword}

\end{frontmatter}


 \section{Introduction}
 \label{}
 By now a substantial amount of work has already been done to understand unexpectedly large
 suppression of single electrons or open heavy mesons coming from the fragmentation of a 
 heavy-quark in nuclear medium.  
 There are two main categories associated with these developments \cite{Majumder:2010qh}: 
 ({\it a}) Calculations that extended radiated energy loss formalism for light flavors to include 
 mass dependent terms, as well as a drag term \cite{Qin:2009gw,Abir:2012pu}. 
 ({\it b}) Calculations that have totally ignored the role of radiative loss and only focussed on drag loss. 
 In both sets of calculations, radiative loss is the results of transverse momentum diffusion 
 experienced by the heavy quark, which is denoted by the jet transport coefficient $\hat{q}$. 
 Till now no calculation of heavy flavor energy loss has investigated the possibility that the 
 drag coefficient $\hat{e}$ (or the longitudinal diffusion coefficient $\hat{e}_2$) 
 may also lead to an additional prominent source of radiative energy loss, beyond that provided by $\hat{q}$. 
 In the higher twist framework, the drag (and longitudinal diffusion) coefficient $\hat{e}$ ($\hat{e}_2$) 
 defined as the loss of light-cone momentum (fluctuation in light-cone momentum) per unit light-cone 
 length as,
 \bea
 \hat{e} = \frac{d  \langle \Delta p^- \rangle }{dL^-},  \,\,\,\,\,\,\,\,\,\,  \hat{e}_2 = \frac{d \langle \Delta {p^-}^2 \rangle}{dL^-} .
 \eea
 Here we assume that the parton is moving in the negative light-cone direction. 
 Though this sub-leading transport coefficients have little effect to the off-shellness of a near 
 on-shell \emph{massless} quark, it has a considerable impact on the off-shellness of a near
 on-shell \emph{massive} quark. In particular this sub-leading transport coefficients will \emph{only} have an effect on the radiative energy loss of a patron where 
 the momentum $p$ is comparable to the mass $M$.  
 %
 %
 We also point out that this mass dependent effect is by no means limited only to the higher-twist scheme, 
 but intrinsic to all other formalisms that have considered the radiative 
 loss from a heavy-quark in a nuclear medium. 
  To delineate the relative importance of these additional new drage terms, we have used power-counting 
  techniques borrowed from Soft-Collinear-Effective-Theory (SCET). 
  This helps one to identify the regime where these mass dependent terms really cause detectable 
  effects on the actual gluon bremsstrahlung spectrum. 
  In this work, being a first attempt, we have considered only the case of single scattering and single emission
  off the propagating heavy quark.




 \section{Deep inelastic scattering and the semi-hard heavy quark}
 We consider the case of deep-inelastic scattering of a hard virtual photon off a 
 hard heavy quark within a large nucleus with mass number $A$ \cite{Abir:2014sxa}.
 We also factorized the propagation of the heavy-quark from the hard 
 scattering vertex which produces the outgoing slow moving heavy-quark. 
 The exchanged virtual photon possesses no transverse momentum, in the Breit frame,  
 \begin{eqnarray}
 q \equiv [q^{+}, q^{-}, q_{\perp}] = \left[q^{+},q^{-},0\right].
 \end{eqnarray} 
 The scattering process under consideration is the following: 
 \begin{eqnarray}
 e(L_1) + A(P) \rightarrow e(L_2) + J_{\cal Q}(L_{\cal Q}) + X .
 \label{chemical_eqn}
 \end{eqnarray}
 As there is no valence heavy-quark within the nucleon, the photon will have to strike a 
 heavy quark from rare ${Q} \bar{Q}$ fluctuations within the sea of partons. 
 In the rest frame of the nucleus we assume that the quark and antiquark are almost at rest 
 $i.e.$ $\left(p_{Q}^{+},p_{Q}^{-}, p_{Q \perp} \right)\sim \left(M/\sqrt{2},~M/\sqrt{2},~0\right)$. 
 %
 %
 %
 Now if the nucleus is boosted by a large boost factor $\gamma$ in the 
 $``+\textquotedblright$ direction, momentum components of the incoming heavy quark will then scales as, 
 \begin{eqnarray}
 p_{\cal Q} = \left[ p_{Q}^{+}, p_{Q}^{-}, p_{Q \perp} \right] \equiv
 \left[ \gamma \frac{M}{\sqrt{2}},~\frac{1}{\gamma}\frac{M}{\sqrt{2}},~0   \right].
 \end{eqnarray}
 Momentum components of the incoming photon have been assumed as, 
 \begin{eqnarray}
 q = \left[ -\gamma \frac{M}{\sqrt{2}} + \frac{M^{2}}{2q^{-}},
 ~ q^{-} - \frac{1}{\gamma}\frac{M}{\sqrt{2}}, ~0   \right] .
 \end{eqnarray}
 As there is a large boost factor $(\gamma)$, one can reasonably assume that $\gamma M \gg M \sim q^{-} \gg 
 M/\gamma $. 
 This would let us to define the hard scale $Q$ as  $Q^2 = - q^2 \simeq \gamma M q^{-}/\sqrt{2}$. Therefore, 
 final out-going quark have momentum components that scales as,  
 \begin{eqnarray}
 p_f=p_{\cal Q} + q = \left[\frac{M^{2}}{2q^{-}} , q^{-} , 0 \right].
 \end{eqnarray}
 \subsection{Power Counting and the small $\lambda$ parameter} 
 In this study, we have introduced the dimensionless small parameter $\lambda$ in order to set up the power counting. 
 We borrowed the concept from soft collinear effective theory (SCET) to represent semi-hard 
 scales as $\lambda Q$ and softer scales as $\lambda^2 Q$. 
 Here we have retained leading and next-to-leading terms in $\lambda$ power counting and neglecting all terms that
 scale with  $\lambda^2$ or a higher power of $\lambda$.
 We have chosen the scaling variable $\lambda$ so that perturbation theory may be applied 
 down to momentum transfer scales even at $\lambda^{3/2}Q\sim \Lambda_{\rm QCD}$.
 In this study,
 \begin{eqnarray}
 1 \gg \sqrt{\lambda} \gg \lambda \gg {\lambda}^{\frac{3}{2}}  .
 \end{eqnarray}
 The virtuality of the hard photon defines the hardest scale in the problem, $Q$.  
 The incoming or initial heavy quark has momentum components $p_i\sim(\lambda^{-\frac{1}{2}},\lambda^{\frac{3}{2}},0)Q$, 
 the outgoing heavy quark has momentum components $p\sim(\sqrt{\lambda},\sqrt{\lambda},0)Q$.
 The mass of the semi-hard heavy quark scales as $M\sim \sqrt{\lambda} Q$. 
 Leading contribution to gluon emission arises from the 
 region where real emitted gluons have momenta which scale as $l \sim(\lambda,\lambda,\lambda)Q$.
 The fraction of light cone momenta carried out by the gluon is $y=l^-/p^- \sim \sqrt{\lambda}$.
 Scaling of the virtual Glauber gluons is as follows, 
 $k\sim(\lambda^{\frac{3}{2}},\lambda^{\frac{3}{2}},\lambda)Q$  with $k^2=2k^+k^--k_\perp^2 \simeq -k_\perp^2$.   
 %
 \section{Induced gluon radiation off the heavy quark}
 \begin{figure}[ht]
   \begin{center}
 \includegraphics[width=7cm,height=3.5cm]{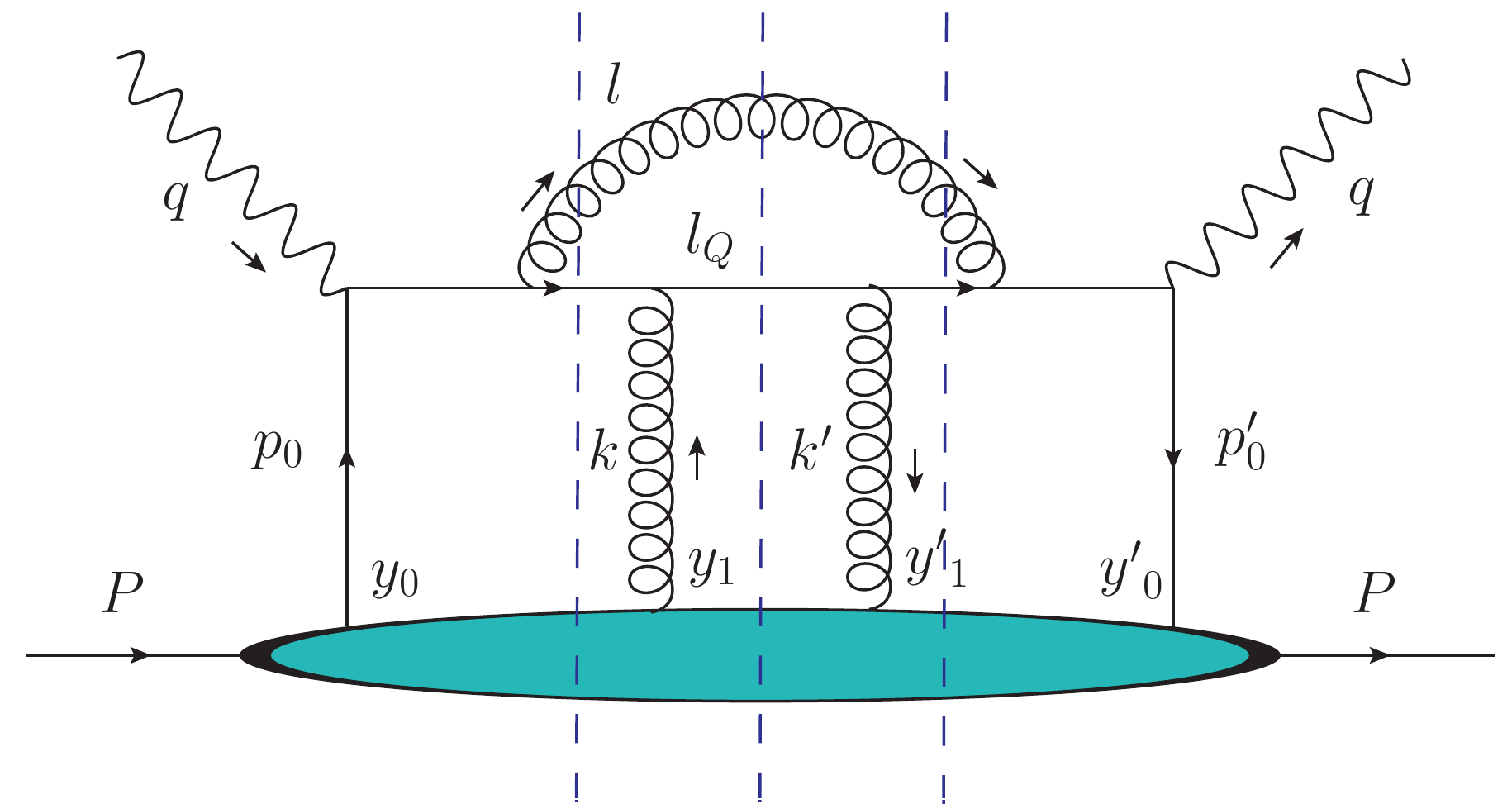}
 \caption{One of the 11 single gluon emission diagram, where gluon emission is induced by 
 single scattering. Dashed lines indicate three separate cuts, denoted as central, left and right.}
      \label{Sample_Diagram}
  \end{center}
 \end{figure}
 In total, there are 11 separate topologically distinct diagrams similar to that in Fig.~\ref{Sample_Diagram} \cite{Zhang:2004qm}. 
 We have defined the following momentum fractions for convenience, 
 %
 \begin{eqnarray}
 y &=& \frac{l^-}{q^-}~,~~
 \eta = \frac{k^-}{l^-} ~,~~   \zeta = \frac{1-y}{1-y+\eta y}~, \nn \\
 x_0 &=& \frac{p_i^+}{P^+}~,~~
 x_1 = \frac{k^+}{P^+}~,~~   \chi=\frac{y^2M^2}{l_\perp^2}~, \nn \\
 x_L &=& \frac{l_\perp^2}{2P^+q^-y(1-y)} ~,~~
 x_D = \frac{k_\perp^2-2l_\perp k_\perp}{2P^+q^-}  ~~,~ \nn \\
 x_K &=& \frac{k_\perp^2}{2P^+q^-} ~,~~\kappa=\frac{1}{1+(1-y)^2}~,~~  \nn \\
 x_M &=& \frac{M^2}{2P^+q^-}. \label{momentum-fractions} 
 \end{eqnarray}
 One obtain the full contribution to the hadronic tensor by adding all the contributions from all the diagrams, 
 We decompose the hadronic tensor as, 
 \bea
 W^{\mu \nu} = g^{4} 2\pi ( - g_{\perp}^{\mu \nu} ) {\cal H}^{c,l,r}.
 \eea
 The entire contribution from all real diagrams is contained in the factor  $\mathcal{H}^{c,l,r}$. In virtual contributions there are 
 no cuts in the final state radiated gluon. We have not considered virtual contributions in this effort. 
 This entire factor ${\cal H}^{c,l,r} $ is obtained as,
 \begin{eqnarray}
  {\cal H}^{c,l,r} &=& \sum_{m,n=1}^{3} {\cal C}_{m,n}^{c,l,r}   
  + \mathcal{C}_{0,1} + \mathcal{C}_{1,0}  \nn \\
 &=& \frac{2\pi \alpha_s}{N_c} \int d l_\perp^2 {H}^{c,l,r} \nn \\
  &\times& \exp\left[ i\left(x_B+x_L+\frac{x_M}{1-y}\right)P^+\left({y'}_0^--{y}_0^-\right)  \right. \nn \\
  &+& i\left(\zeta x_D+\left(\zeta-1\right)\frac{x_M}{1-y}
 -\zeta\frac{\eta y^2}{1-y}x_L\right) \nn \\
&\times& \left. P^+\left({y'}_1^--{y}_1^-\right) \right] \nn \\
&\times&  \langle A | {\bar \psi}(y_0') \gamma^+ A^{+}(y_1')  A^{+}(y_1) \psi(y_0) | A \rangle   
 \end{eqnarray}
 Now one needs to expand the  cross-section in $k_\perp$ and in $k^-$ in order to
 calculate the next-to-leading power contribution to the semi-inclusive  hard partonic cross-section. 
 We then extract the corresponding transport coefficients inside the gluon emission spectrum 
 kernel for the heavy quark.
 All factors of $k_\perp$ and $k^-$ are absorbed as derivatives within the definition of the transport coefficients using integration by parts 
 [e.g. $k_\perp A^+(\vec{y}_\perp) \exp(i \vec{k}_\perp \cdot \vec{y}_\perp) = -i \nabla_\perp A^+(\vec{y}_\perp) 
 \exp(i \vec{k}_\perp \cdot \vec{y}_\perp) \simeq i F^{+\perp} (\vec{y}_\perp)  \exp(i \vec{k}_\perp \cdot \vec{y}_\perp)$]. 
 The four point non-perturbative operator have  also been factored using the usual phenomenological factorization,
 which in case of transverse scattering may be expressed as, 
 \bea
&&  \langle A | {\bar \psi}(y_0') \gamma^+ F^{+}_\perp (y_1')  F^{+}_\perp (y_1) \psi(y_0) | A \rangle   \nn \\
 &\simeq& C^A_{p}  \langle p |   {\bar \psi}(y_0') \gamma^+ \psi(y_0) | p \rangle  \nn \\
 &\times& \frac{\rho}{2p^+} \langle p | F^{+}_\perp (y_1')  F^{+}_\perp (y_1) | p \rangle .
 \nn
 \eea
 The first operator product yield the incoming quark distribution function within one nucleon. 
 The second operator product yield the transport coefficient due to the scattering of the final state, 
 off a gluon. 
 We have assumed the average condition that both nucleons have a momentum $p = P/A$. 
 The nucleon density within the nucleus is denoted by the factor $\rho$, and $C_p^A$ is an 
 overall normalization constant that contain the nucleon density. 
 The factor of $\rho/(2p^+)$ 
 is written separately as that would be absorbed within the definition of the transport coefficient.
  \begin{figure}[ht]
   \begin{center}
 \includegraphics[width=7.5cm,height=5.5cm]{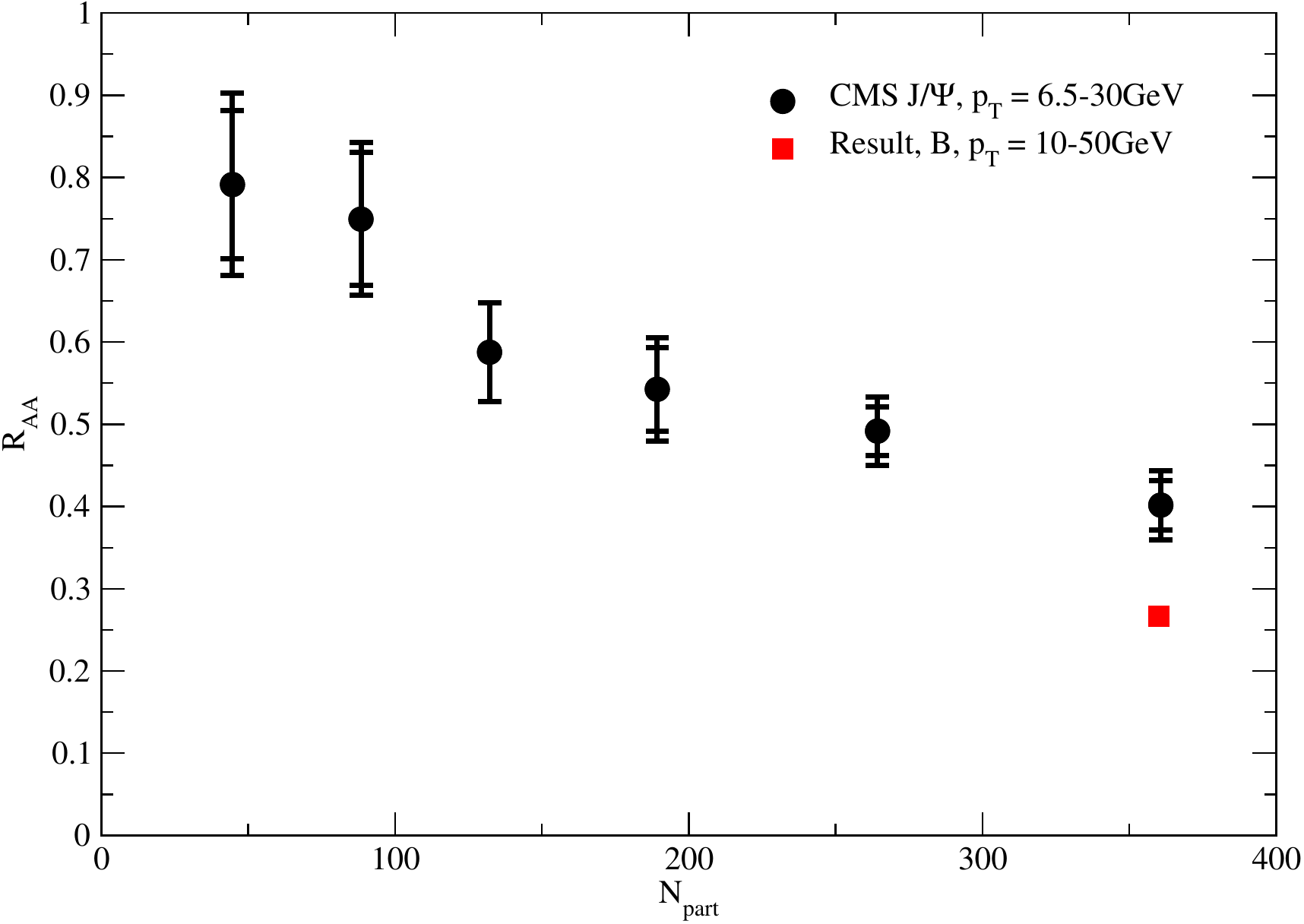}
 \caption{Momentum integrated nuclear modification factor, $R_{AA}$, as function of $N_{\rm part}$.}
      \label{Sample_Diagram_2}
  \end{center}
 \end{figure}
 Expressions for the transverse diffusion $\hat{q}$, the drag (and longitudinal diffusion) coefficient
 $\hat{e}$ ($\hat{e}_2$) can be obtained through derivatives of the kernel with respect to the transverse 
 and $(-)$-light-cone component of the exchange momentum respectively, 
 \begin{eqnarray}
 &~&\left[\nabla_{k_\perp}^2, \nabla_{k^-}, \nabla_{k^-}^2 \right]\left.{H}^{c,l,r}\right|_{k_\perp, k^- = 0}  \nn \\
 &=& 4 C_A \left(\frac{1+(1-y)^2}{y}\right) \frac{l_\perp^4}{[l_\perp^2+y^2M^2]^4} {\tilde H}_{c,l,r}^{{\hat q},{\hat e},{\hat e}_2} .
 \end{eqnarray}
In the equation above, the factor ${\tilde H}_{c,l,r}^{{\hat q},{\hat e},{\hat e}_2} $ represents several terms, depending on the 
cut taken $i.e.$, central $c$, left $l$, or right $r$, and the momentum component with respect to which the Taylor expansion is 
carried out, $i.e.$, $\hat{q}$ for the second derivative in terms of $k_\perp$, $\hat{e}$ for the first derivative with respect to $k^-$  
and $\hat{e}_2$ for the second derivative with respect to $k^-$. Once the derivatives have been 
taken, both factors of the momentum $k_\perp, k^-$ must be set to zero. 
 %
 %
 Finally all the terms can then be combined to obtain the real single gluon emission spectrum. 
 In this work we have retained terms only up to ${\cal O}(\sqrt{\lambda})$, the approximation that has been justified in this study. 
 All terms which scale as ${\cal O}(\lambda)$ or greater, have been neglected. Preliminary estimation of integrated  
 nuclear modification factor both for B and $J/\psi$ with number of participants using our result is rather encouraging, see Fig. [2]. 
 This shows that the gluon bremsstrahlung spectrum from a semi-hard heavy quark 
 is strongly modified by drag induced radiation.
 And the gluon bremsstrahlung spectrum of heavy quark 
 is parametrically sensitive to $\hat e$ which quantifies the amount of drag the moving quark experiences.
 We finally express the gluon spectrum per unit light-cone length as, 

 \begin{eqnarray}
 && \frac{dN_g}{dy dl^2_\perp d\tau}  \nn \\
             &=&2~\frac{\alpha}{\pi}~P(y)~\frac{1}{l^4_\perp} \left(\frac{1}{1+\chi}\right)^{4}
                 ~\sin^2\left(\frac{l^2_\perp}{4l^-(1-y)}(1+\chi)~\tau\right)  \nn \\
          & \times& \left[\left\{\left(1-\frac{y}{2}\right)-\chi
               +\left(1-\frac{y}{2}\right)\chi^2\right\}~{\hat q}
           +\frac{l_\perp^2}{l^-}\chi\left(1+\chi\right)^2~{\hat e} \right.   \nn \\
       && \left.   +\frac{l_\perp^2}{(l^-)^2}\chi\left(\frac{1}{2}-\frac{11}{4}\chi\right){\hat e_2}\right] .   \label{master} 
 \end{eqnarray}
 Three transport coefficients contain the non-perturbative expection of the gluon field strength operators as follows,  
 \begin{eqnarray}
 {\hat q} &=& \varsigma \int \frac{dy^-}{\pi} \frac{\rho}{2p^+}  
           \langle A | F_\perp^{~+}(y^-) F^{\perp+}(0)| A\rangle~e^{-i{\bar \Delta}P^+y^-} , \nn \\
 {\hat e} &=& \varsigma \int \frac{dy^-}{\pi} \frac{\rho}{2p^+}  
         \langle A | i \partial^{-} A^+(y^-) A^+(0)| A\rangle~e^{-i{\bar \Delta}P^+y^-} , \nn \\
 {\hat e_2} &=& \varsigma \int \frac{dy^-}{\pi} \frac{\rho}{2p^+}  
                \langle A | F^{-+}(y^-) F^{-+}(0)| A\rangle~e^{-i{\bar \Delta}P^+y^-} \nn ,
 \end{eqnarray}
 with,
 \begin{eqnarray}
  \varsigma = \frac{4\pi^2 C_R\alpha_s}{N_C^2-1}~.
 \end{eqnarray}
 \noindent In the equations above, one can observe the appearance of a new modified momentum fraction:
 \begin{eqnarray}
 \bar \Delta &=& \zeta x_D+\left(\zeta-1\right)\frac{x_M}{1-y}
 -\zeta\frac{\eta y^2}{1-y}x_L~.
 \end{eqnarray}
 The presence of such a momentum fraction, clearly indicates that the range of momentum fractions in the definition of $\hat{q}$,
$\hat{e}$ and $\hat{e}_2$ for heavy quark scattering is different from that for light flavor energy loss. 
 This also indicates that the actual value of $\hat{q}$, $\hat{e}$ or $\hat{e}_2$ for heavy quarks 
 may be different from that for light quarks.
 Careful analysis of heavy-quark energy loss may then lead to an understanding of the $x$-dependence of the in-medium 
 gluon distribution function that sources transport coefficients. 
 This is essential to understand the degrees of freedom within dense media, where heavy quark loses its energy most.
\section{Conclusion}
 In this work \cite{Abir:2015hta} we have considered a hard virtual photon scattering off a heavy quark (within a proton), 
 that converts it to a slow moving heavy quark which then moves back through the remainder of the nucleus before escaping 
 and fragmenting into a jetwith a leading heavy meson. 
 Here both transverse broadening as well as the longitudinal drag and longitudinal diffusion, have been
 studied on an equal footing.
 We have categorically focussed our study on ``semi-hard'' quarks where the mass and momentum scale as
 $M, p \sim \sqrt{\lambda} Q$, as these are the quarks for which mass modifications is most prominent. 
 Our result can be used to estimate the value of this sub-leading non-perturbative jet transport
 parameter ($\hat e$) from heavy flavor data of heavy-ion collider experiments. 
 These extra additive contributions to the gluon bremsstrahlung spectrum may lead to an 
 eventual solution of the heavy quark puzzle.  
 
 \section{Acknowledgements}
 \label{}
 This work was supported in part by the US National Science Foundation under grant number 
 PHY-1207918 and by the US Department of Energy under grant number DE-SC00013460. This work
 is also supported in part by the Director, Office of Energy Research, Office of High Energy 
 and  Nuclear Physics, Division of Nuclear Physics, of the U.S. Department of Energy, through
 the JET topical collaboration.





\end{document}